\documentclass[11pt]{article}
\usepackage{latexsym}
\usepackage{epsfig}

\def\bg#1{\mbox{\boldmath$#1$}}

\newcommand{\del}{\partial}
\newcommand{\beq}{\begin{eqnarray}}
\newcommand{\eeq}{\end{eqnarray}}
\newcommand{\be}{\begin{eqnarray*}}
\newcommand{\ee}{\end{eqnarray*}}
\newcommand{\bk}{{\bf k}}

\newcommand{\bx}{{\bf x}}

\newcommand{\ra}{\rightarrow}

\newcommand{\ex}[1]{\langle\,#1\,\rangle}

\newcommand{\om}{{\omega}}

\begin{document}

\centerline{\large\bf{Black-body radiation in extra dimensions}}
\vskip 5mm
\centerline{Haavard Alnes\footnote{havard.alnes@fys.uio.no}, Finn Ravndal\footnote{finn.ravndal@fys.uio.no} and Ingunn Kathrine Wehus\footnote{i.k.wehus@fys.uio.no}}
\vskip 5mm
\centerline{\it Institute of Physics, University of Oslo, N-0316 Oslo, Norway.}

\begin{abstract}

\footnotesize{The general form of the Stefan-Boltzmann law for the energy density of black-body radiation is generalized to a spacetime with
extra dimensions using standard kinetic and thermodynamic arguments. From statistical mechanics one obtains an exact formula. 
In a field-theoretic derivation, the Maxwell field must be quantized. The
notion of electric and magnetic fields is different in spacetimes with more than four dimensions. While the energy-momentum tensor
for the Maxwell field is traceless in four dimensions, it is not so when there are extra dimensions. But it is shown that its thermal average is
traceless and in agreement with the thermodynamic results.}

\end{abstract}

\section{Introduction}
Since the introduction of string theory to describe all the fundamental interactions, the possibility that we live in a 
spacetime with more than four dimensions, has steadily generated more interest both in elementary particle physics and nowadays 
also in cosmology. We know that any such extra dimensions must be microscopic if they exist at all and many efforts
are already under way to investigate such a possibility\cite{extra}.

Physical phenomena in spacetimes with extra spatial dimensions will in general be different from what we know in our spacetime of
four dimensions. Although the fundamental laws in such spacetimes can easily be generalized, their manifestations will not be the same.
Even if it turns out that there are no extra dimensions, it is still instructive to investigate what the physics then would be like.

Here we will take a closer look at the Maxwell field and then in particular its properties at finite temperature in the form of
black-body radiation. It is characterized by a pressure $p$ and an energy density $\rho$. In a spacetime with three spatial dimensions, 
these are related by $p = \rho/3$. This can be derived by purely kinematic arguments which are used in the next section to find
the corresponding relation in a $D$-dimensional Minkowski spacetime. When the radiation is in thermal equilibrium at temperature
$T$, we then get by thermodynamic arguments the generalization of the Stefan-Boltzmann law on the form $\rho \propto T^D$.

In the following section these results are derived more accurately using statistical mechanics. One obtains these results by
just assuming that the radiation is composed of massless particles described by quantum mechanics. Except for an overall factor
giving the spin multiplicity, these results should then be the same for scalar particles with no spin, photons which have spin-1 and
gravitons which are the massless spin-2 quanta of gravitation. While our results for the pressure and density corresponds to
a traceless thermal average of the energy-momentum tensor, the general trace of the corresponding energy-momentum tensors of these
fields is not zero. This apparent paradox is discussed and resolved in the last section where we concentrate on the Maxwell field and the
corresponding photons at finite temperature. The quantization of the field in a spacetime with dimensions $D >4$ is a bit more
complicated than in the usual case since the magnetic field can no longer be represented by a vector. Also the number of independent
directions of polarization or helicity states will now be more than two.

\bigskip
{\Large\bf 2\hspace{2mm} Thermodynamics}

In an ordinary spacetime with $D = 3 + 1$ dimensions the pressure of black-body radiation with energy density $\rho$ is given as
$p = \rho/3$. This is most easily derived from a simple kinetic consideration of massless particles impinging on a plane wall 
and thereby being reflected\cite{Gas}. If the angle between the momentum of the incoming particle and the normal to the plane is $\theta$,
then the kinetic pressure is
\beq
           p = \rho\,\ex{\cos^2\theta}                                                                \label{press}
\eeq
where the average is taken over the full spherical angle $2\pi\int_0^\pi d\theta\sin\theta = 4\pi$. In $d=3$ spatial dimensions one
then finds $\ex{\cos^2\theta}=1/3$ which gives the above result for the pressure.

With extra, non-compactified dimensions the pressure will again be given by (\ref{press}). The angular average is then a bit more
cumbersome to evaluate and is worked out in the Appendix. Not so surprising, we then find that the pressure is in general $p = \rho/d$
where $d$ is the number of spatial dimensions.

Assuming that the black-body radiation is described by ordinary thermodynamics, we can now also derive the temperature dependence of
the energy density\cite{Zem}. If the radiation fills a volume $V$ and is in equilibrium with temperature $T$, the total energy $U = \rho V$
will obey the energy equation
\beq
               \left({\del U\over\del V}\right)_T = T \left({\del p\over\del T}\right)_V - p
\eeq
With $p = \rho/d$ this gives
\be
               \rho = {T\over d}{d\rho\over dT} - {\rho\over d}
\ee
which simplifies to
\be
          {d\rho\over\rho} = (d+1){dT\over T}
\ee
One thus obtains for the energy density
\beq
           \rho = CT^D                                           \label{thermo_dens}
\eeq
where $C$ is an integration constant and $D = d+1$ is the dimension of the extended spacetime.

\bigskip
{\Large\bf 3\hspace{2mm} Statistical mechanics}

We will now consider the radiation as made up of massless particles moving in a $d$-dimensional space obeying Bose-Einstein statistics and 
in thermal equilibrium at temperature $T$. It will be convenient to use units so that the speed of light $c=1$ and Planck's constant 
$\hbar = 1$. The energy of one such particle with momentum $\bk$ is then $\omega_\bk = |\bk| = k$ and the internal energy density will be
given by
\beq
        \rho = \int\!{d^dk\over(2\pi)^d} {\omega_\bk\over e^{\beta\omega_\bk} - 1}                       \label{energydens}
\eeq
with  $\beta = 1/(k_BT)$ where $k_B$ is the Boltzmann constant. The differential volume element in momentum space is $d^dk = 
\Omega_{d-1}k^{d-1}dk$ where the full solid angle $\Omega_{d-1}$ is given in the Appendix. Making then use of the integral
\be
           \int_0^\infty\!dx{x^n\over e^x -1 } = \Gamma(n+1)\zeta(n+1)
\ee
where $\zeta(z)$ is Riemann's zeta-function, we have
\beq
           \rho = {\Omega_{d-1}\over(2\pi)^d}\Gamma(D)\zeta(D)(k_BT)^D               \label{Sdense}
\eeq
when expressed in terms of the spacetime dimension $D=d+1$. The temperature dependence is seen to be in agreement with the thermodynamic
result (\ref{thermo_dens}). The pressure of the gas is similarly obtained from the free energy as in the ordinary case for $D=4$. Again one
finds $p = \rho/d$ consistent with the kinetic argument in the first section.

Using the duplication formula for the $\Gamma$-function
\be
            \Gamma(2z) = {2^{2z-{1\over 2}}\over\sqrt{2\pi}}\Gamma\big(z\big)\Gamma\Big({z+{1\over 2}}\Big)\,,
\ee
in the result (\ref{Sdense}) for the density, the formula for the corresponding pressure takes the somewhat simpler form
\beq
                p = {\Gamma(D/2)\over\pi^{D/2}}\zeta(D)(k_BT)^D
\eeq
As a check, we get in the ordinary case of four spacetime dimensions $p = (\pi^2/90)(k_BT)^4$ since $\zeta(4) = \pi^4/90$. This is
essentially the Stefan-Boltzmann law in these particular units. If the massless particles making up the radiation gas has non-zero spin, 
these results must be multiplied by the corresponding spin multiplicity factor.

\bigskip
{\Large\bf 4\hspace{3mm}Scalar quantum field theory}

The massless particles in the previous section will be the quanta of a corresponding quantized field theory. Let us first consider
the simplest case when these are spinless particles described by a scalar field $\phi = \phi(\bx,t)$. The corresponding Lagrangian
is
\beq
            {\cal L} = {1\over 2}\eta^{\mu\nu}\del_\mu\phi\del_\nu\phi \equiv  {1\over 2}(\del_\lambda\phi)^2     \label{SL}
\eeq
where the Greek indices take the values $(0,1,2,\ldots,d)$. The metric in the $D$-dimensional Minkowski spacetime is $\eta_{\mu\nu} =
\mathrm{diag}(1,-1,-1,\ldots,-1)$. From here follows the equation of motion $\del^2\phi = 0$ which is just the massless Klein-Gordon equation.

We are interested in the energy density and pressure of the field. Both follow from the corresponding
canonical energy-momentum tensor\cite{QFT}
\beq
          T_{\mu\nu}=\del_\mu\phi\del_\nu\phi - {1\over 2}\eta_{\mu\nu}(\del_\lambda\phi)^2                  \label{SEM}
\eeq
In a quantized theory it is the expectation value $\ex{T_{\mu\nu}}$  which gives the corresponding measured values. The energy density
is therefore
\beq
          \rho = \ex{T_{00}} = {1\over 2}\ex{{\dot\phi}^2} + {1\over 2}\ex{({\bg\nabla}\phi)^2}             \label{S_dens}
\eeq 
Spherical symmetry implies that the expectation values of all the spatial components are simply given as $\ex{T_{mn}} = p\delta_{mn}$ where
$p$ is the pressure. Thus we find that the pressure is given by the trace
\beq
           p = {1\over d}\ex{T_{nn}} = {1\over d}\ex{({\bg\nabla}\phi)^2} +{1\over 2}\ex{{\dot\phi}^2 - ({\bg\nabla}\phi)^2} \label{Spress}
\eeq
when we use the Einstein convention summing over all equal indices. These expectation values are usually divergent, but the thermal parts 
will be finite for such a free field.

In order to quantize the field, we consider it to be confined to a finite volume $V$ with periodic boundary conditions. It can then be 
expanded in modes which are plane waves with wave vectors $\bk$. The field operator takes the form
\beq
           \phi(\bx,t) = \sum_\bk\sqrt{1\over 2\om_\bk V}\left[a_\bk e^{i(\bk\cdot\bx - \om_\bk t)}
                       + a_\bk^\dagger e^{-i(\bk\cdot\bx - \om_\bk t)}\right]                                 \label{S-op}
\eeq
where $a_\bk$ and  $a_\bk^\dagger$ are annihilation and creation operators with the canonical commutator
\be
         [a_\bk, a_{\bk'}^\dagger] = \delta_{\bk\bk'}
\ee
As in ordinary quantum mechanics, the number of particles in the mode with quantum number $\bk$  is given by the operator 
$a_\bk^\dagger a_\bk$. At finite temperature its expectation value $\ex{a_\bk^\dagger a_\bk}$ equals the Bose-Einstein distribution function
\beq
          n_\bk = {1\over e^{\beta\om_\bk} - 1}                                       \label{BE}
\eeq
This also equals $\ex{a_\bk a_\bk^\dagger}$ when we disregard the vacuum or zero-point contributions which can be neglected at finite 
temperature. Similarly, the expectation values of the operator products $a_\bk a_\bk$ and  the Hermitian adjoint $a_\bk^\dagger a_\bk^\dagger$ 
are zero.

We can now find the energy density from (\ref{S_dens}). For the mode with wavenumber $\bk$ the time derivative ${\dot\phi}$ will pick up
a factor $\om_\bk$ while the gradient ${\bg\nabla}\phi$ will pick up a corresponding factor $\bk$. Thus we get
\be
     \rho =  \sum_\bk{1\over 2\om_\bk V}\big(\om_\bk^2 + \bk^2\big)n_\bk
\ee 
Now letting the volume go to infinity so that
\be
           \sum_\bk \ra V\int\!{d^dk\over(2\pi)^d}
\ee
and using $\bk^2 = \om_\bk^2$ for massless particles, we reproduce the expression (\ref{energydens}) from statistical mechanics. We thus
recover the same result for the thermal energy density. Similarly for the pressure, the two last terms in (\ref{Spress}) will cancel and 
the equation simplifies to $p = \rho/d$ as before.

The expectation value of the full energy-momentum tensor at finite temperature is now diagonal with the components $\ex{T_{\mu\nu}} =
\mathrm{diag}(\rho, p, p,\ldots, p)$. Since the pressure $p = \rho/d$, we have then recovered the well-known fact that the expectation value
of the trace is zero for massless particles, $\ex{T^\mu_{\;\;\mu}} = 0$. What is a bit surprising is that the trace of the energy-momentum 
tensor itself in (\ref{SEM}) is generally not zero. In fact, we have $T^\mu_{\;\;\mu} = (1 - D/2)(\del_\lambda\phi)^2$ which is only zero
in $D=2$ spacetime dimensions. In that case the massless scalar field is said to have conformal invariance\cite{FR}. 
This is a fundamental symmetry in modern string theories. But we know from thermodynamics and statistical mechanics that the expectation 
value of the trace of $T_{\mu\nu}$ is zero in all dimensions. From the above calculation we see that  here in quantum 
field theory, this comes about since we are dealing with massless particles for which $\ex{{\dot\phi}^2} = \ex{({\bg\nabla}\phi)^2}$. This is equivalent to  
$\ex{(\del_\lambda\phi)^2} = 0$ which follows from the equation of motion for the field after a partial integration.

Concerning conformal invariance, the massless scalar field is special. Many years ago Callan, Coleman and Jackiw\cite{CCJ} showed that  
it can be endowed with conformal invariance in any dimension, not only for $D=2$. The corresponding energy-momentum tensor will then be
the canonical one in (\ref{SEM}) plus a new term
\beq
            \Delta T_{\mu\nu} = -{1\over 4}{D-2\over D-1}\Big(\del_\mu\del_\nu - \eta_{\mu\nu}\del^2\Big)\phi^2       \label{hug}
\eeq
Taking now the the trace of this improved energy-momentum tensor, it is  found to be zero when one makes use of the classical field equation 
$\del^2\phi = 0$.  Needless to say, the expectation value of the conformal  piece (\ref{hug}) is zero at finite temperature. 
However, this is not the case at zero temperature for the scalar Casimir energy due to quantum fluctuations between two parallel
plates\cite{S-casimir}. The new term makes the energy density finite between the plates and only when it is included, is the vacuum expectation value of
the trace of the energy-momentum tensor zero. For the scalar field, this can be achieved in any dimension.

\bigskip
{\Large\bf 5\hspace{3mm}Maxwell field and photons}

Black-body radiation is historically considered to be a gas of photons at finite temperature. These massless particles are the quanta of 
the Maxwell field. In a spacetime with $D = d+1$  dimensions, the electromagnetic potential is a vector  $A_\mu(x)$ with $D$ components.
A gauge transformation is defined as $A_\mu \ra A_\mu + \del_\mu\chi$ where $\chi(x)$ is a scalar function. It leaves the Faraday field 
tensor $F_{\mu\nu} = \del_\mu A_\nu - \del_\nu A_\mu$ invariant as in $D=4$ dimensions. Thus the Lagrangian for the field also takes 
the same form,
\beq
           {\cal L} = - {1\over 4}F_{\alpha\beta}F^{\alpha\beta} \equiv  - {1\over 4}F_{\alpha\beta}^2            \label{ML}
\eeq
and is obviously also gauge invariant. While the components $F_{0j}$ form an electric vector ${\bf E}$ with $d$ components, the  $d(d-1)/2$
magnetic components $F_{ij}$ no longer form a vector. The number of electric and magnetic components are equal only when $D=4$.

From the above Lagrangian one gets the corresponding energy-momentum tensor
\beq
      T_{\mu\nu} = F_{\mu\lambda}F^\lambda_{\;\;\nu} + {1\over 4}\eta_{\mu\nu}F_{\alpha\beta}^2              \label{T_EM}
\eeq
as in the ordinary, four-dimensional case\cite{QFT}. The energy density is therefore 
\beq
         T_{00} = E_i^2 + {1\over 4}\Big(F_{ij}^2 - 2E_i^2\Big) = {1\over 2}E_i^2 + {1\over 4}F_{ij}^2          \label{A-dens}
\eeq
while the pressure in the radiation will follow from the spatial components
\beq
        T_{mn}= -E_m E_n + F_{mk} F_{nk} - {1\over 4}\delta_{mn}\Big(F_{ij}^2 - 2E_i^2\Big)                    \label{A-press}
\eeq
In order to calculate the expectation values of these quantities, the field must be quantized. This is most convenient to do in
the Coulomb gauge ${\bg\nabla}\cdot{\bf A} = 0$.  In vacuum, the component $A_0 = 0$ and we have $D-2$  degrees of freedom,
each corresponding to a independent polarization vector ${\bf e}_\lambda$. Corresponding to the expansion of the scalar field operator in
(\ref{S-op}), we now have for the electromagnetic field
\beq
           {\bf A}(\bx,t) = \sum_{\bk\lambda}\sqrt{1\over 2\om_\bk V}\left[{\bf e}_\lambda a_{\bk\lambda} e^{i(\bk\cdot\bx - \om_\bk t)}
                          + {\bf e}_\lambda^* a_{\bk\lambda}^\dagger e^{-i(\bk\cdot\bx - \om_\bk t)}\right]                  \label{A-op}
\eeq
The polarization vectors  for a mode with wavenumber $\bk$ are orthonormalized so that ${\bf e}_\lambda^*\cdot{\bf e}_{\lambda'} 
= \delta_{\lambda\lambda'}$ together with $\bk\cdot{\bf e}_\lambda = 0$ and thus satisfy the completeness relation
\beq
             \sum_\lambda e^*_{\lambda i} e_{\lambda j} = \delta_{ij} - {k_ik_j\over k^2}
\eeq
Since the electric field ${\bf E} = -{\dot{\bf A}}$, we then find that each polarization component has the same expectation value as 
the scalar field in the previous section, i.e.
\beq
                \ex{E_i^2} = (D-2)\sum_\bk{\om_\bk^2\over 2\om_\bk V}n_\bk
\eeq
where $n_\bk$ is the Bose-Einstein density (\ref{BE}). Similarly, for the magnetic components $F_{ij} = \del_iA_j - \del_jA_i$ we obtain
\beq
                \ex{F_{ij}^2} = 2(D-2)\sum_\bk{\bk^2\over 2\om_\bk V}n_\bk
\eeq
when we make use of the above completeness relation for the polarization vectors. Since the photons are massless, we again have $\om_\bk
=|\bk|$ and therefore $\ex{F_{ij}^2} =  2\ex{E_i^2}$. As expected, we then find that the density of the photon gas is  $D-2$ times the 
density of the scalar gas. We also see that the expectation value of the last term in (\ref{A-press}) is zero.  Defining again 
the pressure as $\ex{T_{mn}} = p\delta_{mn}$, it follows then that $p = \rho/d$ as it should be.

Even though the expectation value of the energy-momentum tensor for the electromagnetic field is found to be zero at finite temperature,
the tensor itself in (\ref{T_EM}) has a trace $T^\mu_{\;\;\mu} = (D-4)F_{\mu\nu}^2/4$. It is therefore traceless only in the ordinary
case of $D=4$ spacetime dimensions when the field has conformal invariance. In contrast to the scalar case, the Maxwell field will not have
this symmetry in spacetimes with extra dimensions and there is no corresponding, improved energy-momentum tensor with zero trace.
As we have seen above, this has no serious implications for the photon gas at finite temperature. But at zero temperature when one
calculates the electromagnetic Casimir energy, the tracelessness of $T_{\mu\nu}$ plays an important role\cite{BM}. A similar calculation
of this vacuum energy with extra dimensions would therefore be of interest.

\bigskip
{\Large\bf 6\hspace{3mm}Conclusion}

Black-body radiation is defined to be a free gas of massless particles at finite temperature. Purely kinetic arguments then relates
the pressure and density by $p = \rho/d$ where $d$ is the spatial dimension of the volume containing the gas. This is equivalent to saying 
that the finite-temperature expectation value of the energy-momentum tensor of the gas is zero. 
Describing these particles as quanta of a field theory, we have shown that this tracelessness has very little to do with the trace of the 
corresponding  energy-momentum tensor of the field. 

There are many different ways to derive the energy-momentum tensor. The correct result follows from the most direct derivation starting from
the Lagrangian  ${\cal L}$  of the field in a curved spacetime with metric $g_{\mu\nu}$ instead of the Minkowski metric $\eta_{\mu\nu}$. 
From a variational principle one then finds that the energy-momentum tensor will have the general form\cite{FR}
\beq
          T_{\mu\nu} = 2{\del{\cal L}\over \del g^{\mu\nu}} - g_{\mu\nu}{\cal L}
\eeq
From the scalar Lagrangian (\ref{SL}) this gives the energy-momentum tensor (\ref{SEM}) and similarly for the Maxwell field with
the Lagrangian (\ref{ML}). What we have shown for these fields at finite temperature can be summed up in the statement that
the expectation value $\ex{\cal L}$ in the last term of $T_{\mu\nu}$ is zero. This follows directly from the equations of motion and
is also true for massive fields. The average must be taken over an infinite volume or a finite volume with periodic boundary conditions
so that the surface terms from the partial integrations vanish. 

Here we have only considered the unphysical situation where the extra dimensions are assumed to be infinite in extent like the ones we know. 
In a realistic situation they must be compactified or curled up at a very small scale that is consistent with present-day observations. The
black-body radiation laws derived above will then be modified and end up as small corrections to the four-dimensional results. Such an 
investigation will be more demanding and is not taken up here.

We want to thank Professor R. Jackiw for several useful comments. This work has been supported by grants no. 159637/V30 and 151574/V30 from the
Research Council of Norway.

\bigskip
{\Large\bf Appendix}

In a $d$-dimensional Euclidean space one can specify a point on the unit sphere by $d-1$ angles, $(\phi, \theta_1\theta_2,\ldots,\theta_{d-2})$.
Here $\phi$ is an azimuthal angle with the range $0\le\phi < 2\pi$ while all the polar angles $\theta_n$ vary in the range 
$0\le\theta_n < \pi$. The differential solid angle is then given as
\be
        d\Omega_{d-1} = 2\pi\prod_{n=1}^{d-2}\sin^n\theta_nd\theta_n
\ee 
which gives for the full solid angle in $d$ dimensions 
\be
        \Omega_{d-1} = 2\pi\prod _{n=1}^{d-2}\int_0^\pi\sin^n\theta_nd\theta_n = {2\pi^{d/2}\over\Gamma{(d/2)}}
\ee
when we make use of the integral
\be
            \int_0^\pi d\theta\sin^n\theta = \sqrt{\pi}\,{\Gamma\big({n+1\over 2}\big)\over\Gamma\big({n+2\over 2}\big)}
\ee
For the pressure in black-body radiation we now need the average over this solid angle of $\ex{\cos^2\theta_{d-2}} 
= 1 - \ex{\sin^2\theta_{d-2}}$. With these integral formulas it is now straightforward to show that
\be
       \ex{\sin^2\theta_{d-2}} = {\Gamma\big({d\over 2}\big)\Gamma\big({d+1\over 2}\big)
                             \over\Gamma\big({d-1\over 2}\big)\Gamma\big({d+2\over 2}\big)} = {d-1\over d}
\ee
and thus  $\ex{\cos^2\theta_{d-2}} = 1/d$ as used in the first section of the main text.

\end{document}